\documentclass[aps,reprint,twocolumn,showpacs,prb]{revtex4-1}
\usepackage{graphicx}
\usepackage{dcolumn}
\usepackage{bm}
\usepackage{times}
\usepackage{color}
\begin{document}
\title{Persistence of local-moment antiferromagnetic order in Ba$_{1-x}$K$_x$Mn$_2$As$_2$}
\author{J. Lamsal,$^{1,2}$ G. S. Tucker,$^{1,2}$ T. W. Heitmann,$^{3}$ A. Kreyssig,$^{1,2}$ A. Jesche,$^{1,2}$ Abhishek Pandey,$^{1,2}$ Wei Tian,$^4$ R. J. McQueeney,$^{1,2}$ D. C. Johnston,$^{1,2}$ and A. I. Goldman$^{1,2}$}
\affiliation{$^1$Ames Laboratory, U.S. DOE, Iowa State University, Ames, Iowa 50011, USA}
\affiliation{$^2$Department of Physics and Astronomy, Iowa State University, Ames, Iowa 50011, USA}
\affiliation{$^3$The Missouri Research Reactor, University of Missouri, Columbia, Missouri 65211, USA}
\affiliation{$^4$Quantum Condensed Matter Division, Oak Ridge National Laboratory, Oak Ridge, Tennessee 37831, USA}

\date{\today}

\begin{abstract}
BaMn$_{2}$As$_{2}$ is a local-moment antiferromagnetic insulator with a N\'{e}el temperature of 625 K and a large ordered moment of 3.9 $\mu_{\rm{B}}$/Mn.  Remarkably, this compound can be driven metallic by the substitution of as little as 1.6\%\ K for Ba while retaining essentially the same ordered magnetic moment and N\'{e}el temperature, as previously reported.  Here, using both powder and single crystal neutron diffraction we show that the local moment antiferromagnetic order in Ba$_{1-x}$K$_{x}$Mn$_{2}$As$_{2}$ remains robust up to $x$ = 0.4. The ordered moment is nearly independent of $x$ for 0 $\leq x \leq$ 0.4 and $T_{\rm{N}}$ decreases to 480~K at $x$ = 0.4.
\end{abstract}

\maketitle

The compound BaMn$_2$As$_2$ is of great interest because it is closely related to the parent compounds of the iron pnictide superconductors, but is also a local moment G-type antiferromagnet, similar to parent compounds of the cuprate superconductors. For both the iron pnictides and high-$T_{\rm{c}}$ cuprates, superconductivity appears upon the suppression of long range antiferromagnetic (AFM) order of their parent compounds via chemical substitution or pressure.\cite{Johnston_2010, Paglione_2010, Kimber_2009}  In the iron pnictides, superconductivity is induced by hole or electron doping of the parent system into the metallic state, and the AFM order is widely considered to be best characterized as a spin density wave arising from itinerant electron carriers.\cite{Johnston_2010,Pratt_2011} In the cuprates, superconductivity is induced by electron or hole doping into the Mott-Hubbard type insulating phase of the parent system and the AFM order arises from local moments.  Since BaMn$_2$As$_2$ shares characteristics of both classes of high-$T_{\rm{c}}$ superconducting families it represents a bridge between these families that may shed further light on the nature of superconductivity in these compounds.\cite{Singh_2009}

Electrical resistivity and specific heat measurements on powders and flux-grown single crystals of BaMn$_{2}$As$_{2}$ have demonstrated that this compound is a small-band-gap ($E_{\rm{gap}} \approx 0.05$ eV) semiconductor with an electronic linear specific heat coefficient $\gamma$ = 0, consistent with an insulating ground state.\cite{Singh_2009b} Below $T_{\rm{N}}$ = 625 K, BaMn$_{2}$As$_{2}$ orders into a G-type AFM structure (a collinear AFM structure in which nearest-neighbor Mn moments in the tetragonal basal plane are antiparallel and successive planes along the $c$ axis are also antiferromagnetically aligned) with an ordered moment at 10 K of $\mu$ = 3.88(4) $\mu_{\rm{B}}$/Mn aligned along the tetragonal $c$ axis.\cite{Singh_2009} Further extensive experimental and theoretical studies established that the Mn$^{2+}$ ions possess local moments with spin $S = \frac{5}{2}$.\cite{Johnston_2011} Because of the substantial ordered local moment, as compared to the smaller ordered itinerant moment ($\sim$ 1 $\mu_{\rm{B}}$/Fe) characteristic of the iron pnictides, it was suggested that chemical substitutions might suppress the AFM order and induce large magnetic fluctuations leading to high values of $T_{\rm{c}}$.\cite{Johnston_2010,Simonson_2012,Sun_2012}  Unfortunately, transition metal substitutions on the Mn site and Sb substitution for As either show limited solubility or, in the case of Cr and Fe, do not alter the properties of the parent phase in any significant way.\cite{Pandey_2011,Pandey_2012}

It was recently demonstrated, however, that K substitution for Ba, or the application of external pressure, dramatically alters the electronic properties.\cite{Pandey_2012,Satya_2011,Satya_2011} The substitution of as little as 1.6\%\ of K for Ba in Ba$_{1-x}$K$_{x}$Mn$_{2}$As$_{2}$ results in a metallic ground state.\cite{Pandey_2012}  A similar transition to a metallic state was observed at an applied pressure of 4.5 GPa at 36 K, or $\sim$ 5.8 GPa at 300 K in undoped BaMn$_2$As$_2$.\cite{Satya_2011}  Furthermore, at least for K substitution, both magnetic susceptibility and magnetic neutron diffraction measurements demonstrated that the local moment AFM order is robust since $T_{\rm{N}}$ and the ordered moment of the $x$ = 0.05 K-substituted sample changed only slightly [$T_{\rm{N}}$= 607(2) K and $\mu$ = 4.21(12)$\mu_{\rm{B}}$/Mn] from the values obtained for the parent compound.  Recent studies of more heavily K-doped crystals ($x = 0.19$ and $x = 0.26$) have indicated the presence of weak ferromagnetism below approximately 50 K, with small low-$T$ saturation moments of 0.02 -- 0.08 $\mu_{\rm{B}}$/Mn aligned in the $ab$-plane, that coexists with the Mn AFM order.\cite{Bao_2012} Our present neutron measurements are not sensitive to this small ferromagnetic component.

Two important issues regarding the magnetism in the K-substituted BaMn$_{2}$As$_{2}$ are addressed in this Rapid Communication via neutron diffraction measurements on polycrystalline and single-crystal samples.  First, we find that the G-type AFM order, with the moment directed along the $c$ axis, is very robust for $x$ up to our limit of 0.4, and the magnitude of the ordered moment per Mn, $\mu$, is nearly independent of $x$.  However, $T_{\rm{N}}$ decreases by $\approx140$~K at $x$ = 0.4.  These results contrast with the corresponding behaviors observed for the $A$Fe$_2$As$_2$-type ($A$ = Ba, Sr, Ca) compounds where $T_{\rm{N}}$ and $\mu$ are both suppressed for even small elemental substitutions.\cite{Johnston_2010}

Polycrystalline samples of Ba$_{1-x}$K$_x$Mn$_2$As$_2$ were synthesized by solid-state reaction using Ba (99.99\%) from Sigma Aldrich, and K (99.95\%), Mn (99.95\%) and As (99.99999\%) from Alfa Aesar. Stoichiometric mixtures of the elements were sealed inside Ta tubes that were then sealed in evacuated quartz tubes. The samples were heated to 585\,$^\circ$C at a rate of 40\,$^\circ$C/h, held there for 20\,h, then heated to 620\,$^\circ$C at a rate of 7\,$^\circ$C/h, held there for 20\,h, and finally heated to 750\,$^\circ$C at 40\,$^\circ$C/h and held there for 30\,h. After furnace cooling, the samples were thoroughly ground, pelletized and again sealed in Ta tubes which were subsequenty sealed in evacuated quartz tubes. The samples were heated at 850\,$^\circ$C and then at 1100\,$^\circ$C for 25~h and 65~h, respectively. Then the temperature was raised to 1150\,$^\circ$C and held there for 5~h, and the samples were then quenched in liquid nitrogen.

Single crystals of Ba$_{0.6}$K$_{0.4}$Mn$_2$As$_2$ were grown using MnAs flux. The starting materials Ba, K and MnAs were taken in a ratio of 0.2\,:\,0.8\,:\,5 and placed in a 2~mL alumina crucible which was sealed inside a 5/8 inch diameter Ta tube. The Ta tube containing the sample was sealed inside a quartz tube filled with argon at $\approx 1/4$~atm.~pressure, heated to 700\,$^\circ$C at 100\,$^\circ$C/h, held there for 10\,h, heated to 1230\,$^\circ$C at 50\,$^\circ$C/h and held there for 5\,h and finally cooled to 1110\,$^\circ$C in 70\,h. At this temperature the excess flux was decanted using a centrifuge and shiny plate-like crystals of typical size $4\times 4\times 0.2$~mm$^3$ were obtained. The chemical composition of the crystals was determined using wavelength dispersive x-ray spectroscopy in a JEOL JXA-8200 Superprobe electron probe microanalyzer to be \rm{Ba$_{0.60\mp0.02}$K$_{0.40\pm0.02}$Mn$_2$As$_2$}. This chemical analysis showed that the potassium concentration in the crystals was substantially smaller than the starting ratio ${\rm Ba_{0.2}K_{0.8}}$.

\begin{figure}
\centering\includegraphics[width=0.9\linewidth]{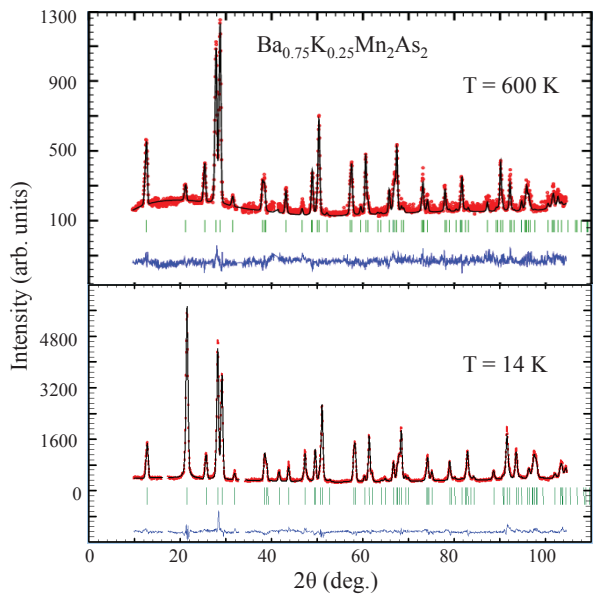}\\
\caption{(Color online) Neutron diffraction profile fits from \mbox{Rietveld} analysis for nuclear and magnetic Bragg reflections from Ba$_{0.75}$K$_{0.25}$Mn$_2$As$_2$ at 14 K and 600 K.  The observed data points are given by red circles, the calculated intensity pattern by black solid lines and corresponding residual (i.e., the difference between observed and calculated pattern) by the blue curve at the bottom. Small angular regions that contain peaks from the MnO second phase were excluded in the refinement.  The green vertical upper tick marks reflect the Bragg positions for nuclear reflections and lower green tick marks give the Bragg positions for magnetic reflections.}\label{Fig_1}
\end{figure}

Neutron powder diffraction measurements were performed on the high-resolution powder diffractometer  at the University of Missouri Research Reactor using a double focusing Si (511) crystal monochromator to select neutrons with a wavelength $\lambda$ = 1.4803~\AA.\cite{Yelon_1998} The monochromator is optimized for the use of small samples, and a position-sensitive detector collects a series of 19$^{\circ}$ sections of the diffraction pattern; the full diffraction pattern was measured in five steps (2$\theta_{\rm{max}}$ = 108$^{\circ}$). The powder diffractometer uses a radial oscillating collimator averaging the shadow of the collimator blades at every channel. The powder samples were loaded into vanadium holders and diffraction patterns were taken between 14 K and 300 K using a closed-cycle refrigerator, and between 300 K and 650 K using a high-temperature furnace. Analysis of the neutron diffraction data was performed by the Rietveld method using FULLPROF.\cite{Carvajal_1993}

\begin{figure}
\centering\includegraphics[width=0.9\linewidth]{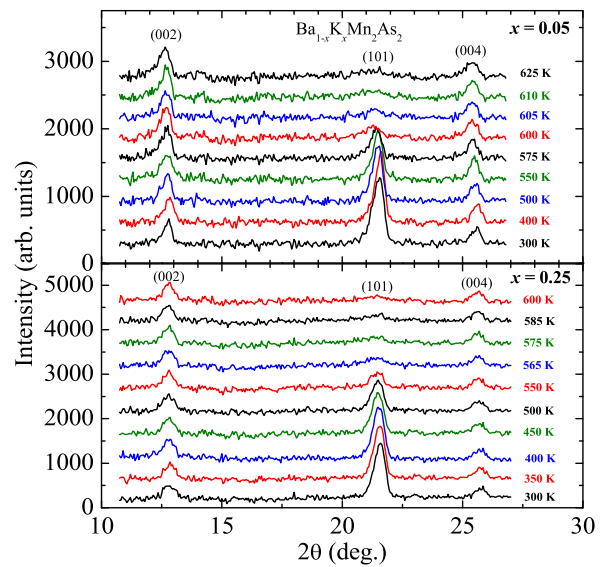}\\
\caption{(Color online) Neutron diffraction data collected in the first segment of the powder diffraction pattern from Ba$_{1-x}$K$_x$Mn$_2$As$_2$ ($x$ = 0.05 and 0.25) in the high-temperature furnace. The background from the empty furnace, measured at 300~K, has been subtracted from the data at each temperature.}\label{Fig_2}
\end{figure}

Single-crystal neutron diffraction measurements were performed on a 27 mg sample of Ba$_{0.6}$K$_{0.4}$Mn$_{2}$As$_{2}$ using the HB1A triple-axis spectrometer at the High Flux Isotope Reactor at Oak Ridge National Laboratory. The experiment was performed with collimations of 48$^\prime$- 40$^\prime$- 40$^\prime$-136$^\prime$, a fixed incident neutron energy of 14.6 meV, and two pyrolytic graphite filters for the elimination of higher harmonics in the incident beam.  The single crystal was mounted in a high-temperature displex cryostat (30 K -- 700 K) with the ($H 0 L$) reciprocal lattice plane coincident with the scattering plane.

\begin{table*}[tbp]
\caption{\label{Table:CrystalData}  Results of the profile refinements from the powder ($x$ = 0, 0.05, 0.25) and single-crystal ($x$ = 0.4) neutron diffraction measurements. Listed are the $a$ and $c$ lattice constants, the unit cell volume ($V_{\rm cell}$), the position parameter for the arsenic atoms ($z_{\rm As}$), the N\'{e}el Temperature ($T_{\rm N}$) and low-temperature ordered magnetic moment ($\mu$).  Errors are given as one standard deviation.}
\begin{ruledtabular}
		\begin{tabular}{l c c c c c c c c}
		 Compound       &   $a$  &  $c$  & $V_{\rm cell}$ & $z_{\rm As}$ & $T_{\rm N}$ &     $\mu$       & Remarks \\
	                    &  (\AA) & (\AA) &  (\AA$^3$)     &              &     (K)     & ($\mu_{\rm B}$) &          \\
		  \hline

            ${\rm BaMn_2As_2}$ & 4.1539(2) & 13.4149(8) & 231.47(2) & 0.3613(2) & 625(1) & 3.88(4) & T = 10 K, Ref.~\onlinecite{Singh_2009} \\
			
            ${\rm Ba_{0.95}K_{0.05}Mn_2As_{2}}$ & 4.1563(4) & 13.4046(13) & 232.57(4)  & 0.3642(5) & 607(2) & 4.21(12) & T = 14 K \\

            ${\rm Ba_{0.75}K_{0.25}Mn_2As_{2}}$ & 4.1569(3) & 13.3072(12) & 229.95(4)  & 0.3621(4) & 575(4) & 4.25(12) & T = 14 K \\

            ${\rm Ba_{0.6}K_{0.4}Mn_2As_{2}}$ & - & - & -  & - & 480(2) & 3.85(15) & T = 30 K, single crystal \\

		\end{tabular}
\end{ruledtabular}
\end{table*}

Examples of the powder profile refinement at 600 K (above $T_{\rm{N}}$) and at our base temperature of 14 K for polycrystalline Ba$_{1-x}$K$_{x}$Mn$_{2}$As$_{2}$ ($x$ = 0.25) are shown in Fig.~\ref{Fig_1}. The Rietveld analysis of the neutron diffraction data showed that both the $x$ = 0.05 and $x$ = 0.25 polycrystalline samples retain the $I4/mmm$ space group over the entire temperature range of these measurements (see Table~\ref{Table:CrystalData}). No second phase, at a detection limit of 1.5 wt.\%, was found for the $x$ = 0.05 sample although approximately 2.2 wt.\%\ MnO was detected as a second phase for $x$ = 0.25. The refined base temperature lattice parameters for $x$ = 0.05 and 0.25 demonstrate the contraction of the $c$-axis lattice parameter with respect to BaMn$_{2}$As$_{2}$, measured at 10 K, consistent with the introduction of the smaller K ion on the Ba site.

The magnetic structure for both the $x$ = 0.05 and 0.25 polycrystalline samples at $T$ = 14 K were refined in the space group $P\overline{1}$ allowing for the possibility of complex magnetic structures.  However, only the G-type AFM structure was necessary to fit the powder diffraction profiles. For this AFM structure magnetic Bragg peaks are found at reciprocal lattice points ($H 0 L$) with $H$ and $L$ odd, coincident with the allowed nuclear reflections.

\begin{figure}[h!]
\centering\includegraphics[width=0.9\linewidth]{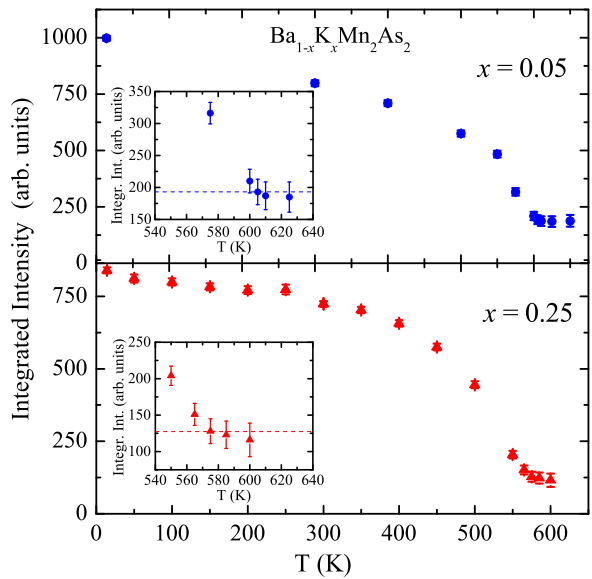}\\
\caption{(Color online) Integrated intensity of the (1~0~1) magnetic Bragg reflection measured for Ba$_{1-x}$K$_x$Mn$_2$As$_2$ ($x$ = 0.05 and 0.25).  Note that there is a low intensity nuclear Bragg peak at (1~0~1) present at all temperatures for both samples.  The insets show expanded plots near $T_{\rm{N}}$ for both samples.}\label{Fig_3}
\end{figure}

To monitor the temperature dependence of the AFM order we followed the evolution of the diffraction pattern in the vicinity of the (1 0 1) diffraction peak which manifests small nuclear and large magnetic contributions.  The data for both samples are summarized in Figs.~\ref{Fig_2} and \ref{Fig_3}, establishing $T_{\rm{N}}$ = 607(2)~K and 575(4)~K for the $x$ = 0.05 and 0.25 samples, respectively.  The refined values for the ordered moments for $x$ = 0.05 and 0.25 were 4.21(12) $\mu_{\rm{B}}$/Mn and 4.25(12) $\mu_{\rm{B}}$/Mn, respectively, showing little change or, perhaps, a small increase at these doping concentrations relative to the ordered moment previously determined for the parent BaMn$_2$As$_2$ compound of 3.88(4) $\mu_{\rm{B}}$/Mn.\cite{Singh_2009}

\begin{figure}[t!]
\centering\includegraphics[width=0.9\linewidth]{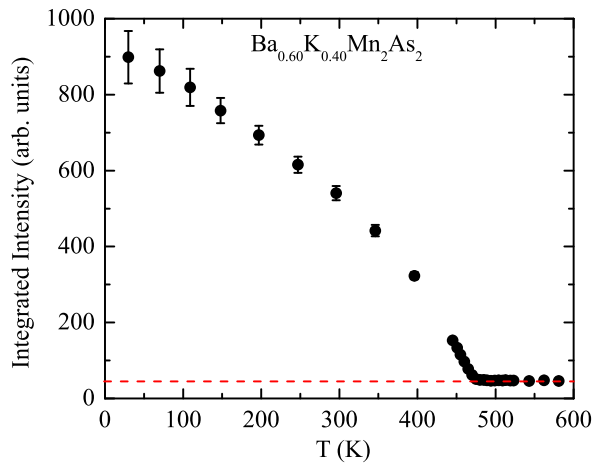}\\
\caption{(Color online) Integrated intensity of the (1 0 1) Bragg peak from the Ba$_{0.6}$K$_{0.4}$Mn$_{2}$As$_{2}$ single-crystal sample.  The dashed horizontal red line represents the contribution of the nuclear scattering to the Bragg peak intensity that is present at all temperatures.}\label{Fig_4}
\end{figure}

Fig.~\ref{Fig_4} shows the temperature dependence of the integrated intensity of the (1 0 1) magnetic and nuclear Bragg peak for the single-crystal Ba$_{0.6}$K$_{0.4}$Mn$_{2}$As$_{2}$ sample. The residual intensity above $T_{\rm{N}}$ = 480(2) K is, again, the nuclear contribution to the Bragg scattering whereas the increase in intensity below $T_{\rm{N}}$ arises from the magnetic ordering in the G-type AFM structure.  Using the (0 0 4) and (2 0 0) nuclear peaks for reference, the ordered moment at $T$ = 30 K for the Ba$_{0.6}$K$_{0.4}$Mn$_{2}$As$_{2}$ sample was determined to be 3.85(15) $\mu_{\rm{B}}$/Mn. The magnetic order parameter evolves smoothly as temperature decreases and does not show any clear evidence of an anomaly at low temperature that might signal a canting of the moment away from the $c$-axis to produce a ferromagnetic component. Furthermore, no anomalies in the intensities of the Bragg peaks arising from the chemical structure were observed at low temperature, although some additional diffraction peak broadening was found below $\sim$ 100 K.  Experiments are planned to further investigate the proposed weak ferromagnetism\cite{Bao_2012} at low temperature.

\begin{figure}[t!]
\centering\includegraphics[width=0.9\linewidth]{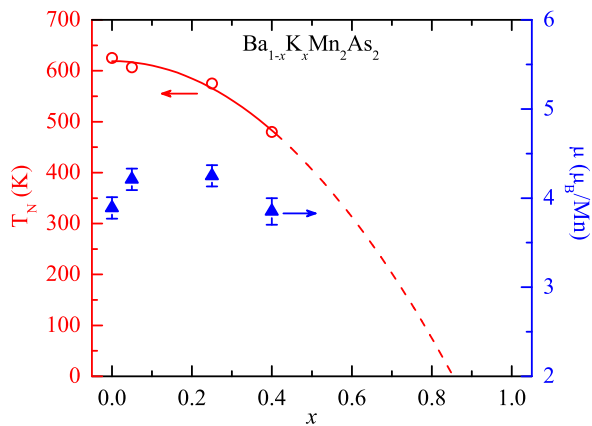}\\
\caption{(Color online) Summary of results for $T_{\rm{N}}$ (open circles) and the ordered moment (solid triangles) for Ba$_{1-x}$K$_x$Mn$_2$As$_2$ as a function of K content.  The solid (dashed) line represents a fit (extrapolation) of the data using a second order polynomial in $x$.}\label{Fig_5}
\end{figure}

A summary of the results for the four samples studied by neutron diffraction is presented in Table~\ref{Table:CrystalData} and in Fig.~\ref{Fig_5} where we plot both $T_{\rm{N}}$ and the low-temperature ordered moment, $\mu$, as a function of K concentration. The N\'{e}el temperature decreases only slowly over the entire range of compositions studied. For a rough estimate of the composition ($x_{\rm{c}}$) where magnetic order may be fully suppressed, we assume a smooth variation in $T_{\rm{N}}$ with K-concentration.  We found that a second-order polynomial fit to the the data for $x \leq 0.4$ (solid line) provides a good description of the observed trend in $T_{\rm{N}}$ versus $x$ and, extrapolating the fit to higher K concentrations (dashed line), we obtain $x_{\rm{c}} \sim 0.85$. The maximum K concentration we have obtained so far is $x \approx 0.5$.

We also see from Fig.~\ref{Fig_5} that the magnitude of the ordered moment remains nearly constant over the composition range of these measurements, although there seems to be some tendency for a slight increase at the lower doping concentrations.  Indeed, it has previously been suggested that a small increase in the ordered moment may result from a reduction in the spin-dependent hybridization between the Mn $3d$ and As $4p$ states with K substitution.\cite{An_2009}  This is also supported by the electronic structure calculations of Bao \emph{et al}., who note that K-substitution weakens the \emph{p-d} hybridization by reducing the As-4\emph{p} weight at the Fermi energy.\cite{Bao_2012}

It is useful to place this work on Ba$_{1-x}$K$_x$Mn$_2$As$_2$ within the context of other recent investigations of related compounds.  For example, a metallic state can also be obtained in BaMn$_2$As$_2$ under a pressure of $\sim$ 5.8 GPa.\cite{Satya_2011}  The insulator-to-metal transition in these measurements, however, is also associated with anomalies in the pressure dependence of the lattice parameters and unit cell volume.  The authors further found a sharp drop in the sample resistance at 17 K, suggesting a superconducting transition.  However, it remains an open question, for further study, whether the magnetic properties of BaMn$_2$As$_2$ under applied pressure are similar to those of Ba$_{1-x}$K$_x$Mn$_2$As$_2$ studied here.

Insulator-to-metal transitions with doping and/or under pressure have also been found in compounds that are closely related to the 1111-type iron arsenides, including LaMnPO (Ref.~\onlinecite{Simonson_2012}) and LaMnAsO.\cite{Sun_2012}  Similar to the case of Ba$_{1-x}$K$_x$Mn$_2$As$_2$, the AFM order in F-doped LaMnPO$_{1-x}$F$_x$ is robust for electron doping up to rather high F concentrations, but no insulator-to-metal transition is realized for $x \leq 0.3$.  Rather, the authors of Ref.~\onlinecite{Simonson_2012} suggest that electron-doping results in the formation of states within the insulating gap that may be regarded as a precursor of the insulator-metal transition found for Ba$_{1-x}$K$_x$Mn$_2$As$_2$. Interestingly, their electronic structure calculations predict a collapse of the insulating gap under applied pressure in two steps: the first to a metallic antiferromagnet with a reduced moment followed by a quenching of the Mn moment at higher pressure as the Mn-P distance decreases below a critical value.\cite{Simonson_2012}  An insulator-to-metal transition was found for Sr-doped La$_{1-x}$Sr$_x$MnAsO at $x \approx 0.08$ and, similar to what we find for Ba$_{1-x}$K$_x$Mn$_2$As$_2$, the N\'{e}el temperature is only gradually suppressed with increasing Sr doping.  It is not clear from these measurements, however, how the ordered magnetic moment changes with Sr-doping in LaMnAsO.\cite{Sun_2012}

In summary, it was previously found that the local-moment AFM ordering associated with Mn is preserved in Ba$_{1-x}$K$_x$Mn$_2$As$_2$ up to $x = 0.05$, even as the compound changes dramatically from a small-gap semiconductor to a metal.\cite{Pandey_2012}  We have shown here that this local-moment behavior is very robust for K concentrations up to, at least, 40\%.  This provides strong evidence that the magnetic exchange between the Mn spins is relatively insensitive to the addition of charge carriers via chemical substitution for Ba, and highlights the local-moment nature of AFM in BaMn$_2$As$_2$ as opposed to the itinerant spin-density wave character previously found for the Ba$_{1-x}$K$_x$Fe$_2$As$_2$ superconductor.\cite{Avci_2012,Dai_2012}

\begin{acknowledgments}
The authors gratefully acknowledge useful discussions with M. Ramazanoglu. Work at the Ames Laboratory was supported by the Division of Materials Sciences and Engineering, Office of Basic Energy Sciences, U.S. Department of Energy, under Contract No. DE-AC02-07CH11358. Work at the High Flux Isotope Reactor, Oak Ridge National Laboratory, was sponsored by the Scientific User Facilities Division, Office of Basic Energy Sciences, U.S. Department of Energy.
\end{acknowledgments}


\end{document}